\documentclass[journal,twoside,web]{ieeecolor}
\usepackage{tmi}
\usepackage{cite}
\usepackage{amsmath,amssymb,amsfonts}
\usepackage{booktabs}
\usepackage{multirow} 
\usepackage{algorithmic}
\usepackage{graphicx}
\usepackage{textcomp}
\usepackage{makecell}
\def\BibTeX{{\rm B\kern-.05em{\sc i\kern-.025em b}\kern-.08em
    T\kern-.1667em\lower.7ex\hbox{E}\kern-.125emX}}
\begin{document}
\title{Learning Robust Medical Image Segmentation from Multi-source Annotations}
\author{Yifeng Wang, Luyang Luo, \IEEEmembership{Member, IEEE}, Mingxiang Wu, Qiong Wang, and Hao Chen, \IEEEmembership{Senior Member, IEEE}
\thanks{This work was supported by Shenzhen Science and Technology Innovation Committee Fund (Project No. SGDX20210823103201011) and in part by the Project of Hetao Shenzhen-Hong Kong Science and Technology Innovation Cooperation Zone (HZQB-KCZYB-2020083).}
\thanks{Yifeng Wang is with the Shenzhen International Graduate School, Tsinghua University, Shenzhen, China (e-mail:wangyife21@mails.tsinghua.edu.cn). }
\thanks{Luyang Luo and Hao Chen are with Department of Computer Science and Engineering, Hong Kong University of Science and Technology, Hong Kong, China. (e-mails:cseluyang@ust.hk, jhc@cse.ust.hk}
\thanks{Mingxiang Wu is with Shenzhen People's Hospital, Shenzhen, China. (e-mail:szmxwu@outlook.com). }
\thanks{Qiong Wang is with Shenzhen Institutes of Advanced Technology, Chinese Academy of Sciences, Shenzhen, China. (e-mail: wangqiong@siat.ac.cn). }
\thanks{Hao Chen is also with HKUST Shenzhen-Hong Kong Collaborative Innovation Research Institute, Futian, Shenzhen, China.). }
\thanks{Yifeng Wang and Luyang Luo contributed equally to this work. }
\thanks{Corresponding author: Hao Chen.}
\thanks{The source code will be released.}
}

\maketitle

\begin{abstract}
Collecting annotations from multiple independent sources could mitigate the impact of potential noises and biases from a single source, which is a common practice in medical image segmentation.
Learning segmentation networks from multi-source annotations remains a challenge due to the uncertainties brought by the variance of annotations and the quality of images.
In this paper, we propose an Uncertainty-guided Multi-source Annotation Network (UMA-Net), which guides the training process by uncertainty estimation at both the pixel and the image levels. 
First, we developed the annotation uncertainty estimation module (AUEM) to learn the pixel-wise uncertainty of each annotation, which then guided the network to learn from reliable pixels by weighted segmentation loss. 
Second, a quality assessment module (QAM) was proposed to assess the image-level quality of the input samples based on the former assessed annotation uncertainties. Importantly, we introduced an auxiliary predictor to learn from the low-quality samples instead of discarding them, which ensured the preservation of their representation knowledge in the backbone without directly accumulating errors within the primary predictor.
Extensive experiments demonstrated the effectiveness and feasibility of our proposed UMA-Net on various datasets, including 2D chest X-ray segmentation, fundus image segmentation, and 3D breast DCE-MRI segmentation.
\end{abstract}




\begin{figure}[t]
  \centering
  \includegraphics[width=\linewidth]{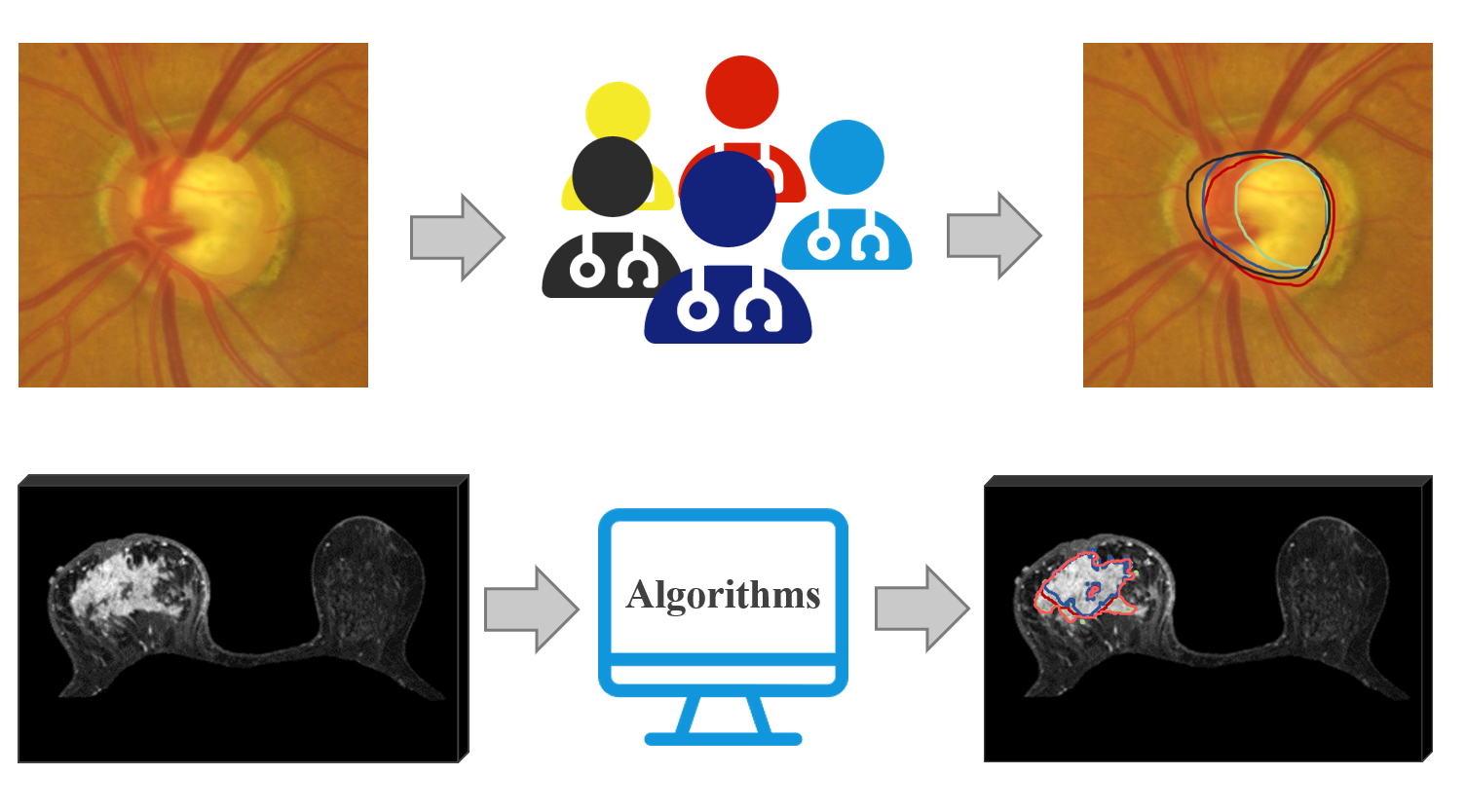}
  \caption{Multi-source annotations could mitigate the noises and biases by a single source. Specifically, the annotations can be acquired by multiple raters (top) or different algorithms when human annotations are not available (bottom).}
  \label{f1}
\end{figure}

\section{Introduction}
\label{sec:introduction}
\IEEEPARstart{M}{edical} image segmentation plays an important role in computer-aided diagnosis\cite{litjens2017survey}. Taking advantage of Convolutional Neural Networks (CNNs), automated segmentation systems have achieved promising performance in the past decade \cite{litjens2017survey}. Currently, training advanced CNNs often requires images paired with well-annotated, single-source annotations. 
However, due to the several reasons such as the raters' individual preferences, limited expertise level, or systematic biases, single-source annotations may contain noises that can undermine the quality of the network. 
To exacerbate the matter, CNNs tend to overfit the noise in annotations from a single source \cite{feng2022learning}. 
A common practice to tackle this challenge is collecting annotations from multiple sources \cite{almazroa2017agreement,orlando2020refuge,armato2011lung, nguyen2019deepusps,wang2022multi}, such as different raters or various automated segmentation algorithms, as shown in Fig. \ref{f1}


Due to differences in the expertise level and individual preferences of different annotators, their annotations of the same medical image may vary greatly. 
To learn from the diverse multi-source annotations, label fusion strategies,such as majority vote and STAPLE \cite{warfield2004simultaneous}, are often used to obtain the ground truth by aggregating the multiple annotations into a unique annotation. 
However, the inter-observer variability information, which represents the commonality and variety among multi-source annotations, is overlooked during the fusion process \cite{ji2021learning}.
Importantly, the commonality and variety of these annotations contain rich information that has the potential to assist in calibrating models to learn from more reliable samples.

To date, several methods have been developed to leverage the variability information among the multi-source annotations \cite{jensen2019improving,jungo2018effect,ji2021learning,wu2022learning,zhang2023multi}. 
For instance, MA-Net \cite{ji2021learning} embedded an expertise-aware inferring module into UNet to adapt to the different expertness of raters. 
Self-Calib \cite{wu2022learning} attempted to find the latent ground truth by recurrently running a diverging and converging model and estimating the expertness map of each annotation. 
These methods were more capable of leveraging and commonality and variety among multiple annotations and showed better performance than the fusing strategies. 
However, these methods have overlooked the proper handling of low-quality samples that surpass the capabilities of the calibration framework, leading to a degradation in the network's performance.
Besides, they often need extra modules or auxiliary branches during the inference stage, which increases the computational burden.

In this paper, we propose Uncertainty-guided Multi-source Annotation Network (UMA-Net) to enable image-level and pixel-level guidance of the network to learn from reliable pixels and high-quality samples. 
First, for the pixel-level guidance, we developed an Annotation Uncertainty Estimation Module (AUEM) to generate uncertainty maps for each annotation, which guides the segmentation network to learn from more reliable pixel labels by weighted segmentation loss. 
We then derived the calibration map from the noisy annotations and their corresponding uncertainty maps, 
which represented a coarse estimation of the latent ground truth.
A consistency regularization was injected among these calibration maps to strengthen the cross-referencing ability of AUEM and minimize the influence of the annotations variation. 
Further, for the image-level guidance, we designed the Quality Assessment Module (QAM) to categorize different samples into high-quality and low-quality groups based on the image-level uncertainty estimation.
Then, we implemented a primary predictor head and an auxiliary predictor head onto the segmentation network to learn from the high-quality samples and the low-quality samples, respectively.
In this way, the representation knowledge of the low-quality samples could be preserved without sacrificing the accuracy of the primary predictor.

Combining AUEM and QAM, we manage to guide the learning of the segmentation network with uncertainty guidance from both the pixel and the image levels.
Moreover, both modules can be discarded freely during testing, and only the segmentation backbone with the primary predictor are reserved, which saved the computational resources.
We conducted extensive experiments with various datasets, including 2D chest X-rays, fundus images, as well as 3D breast DCE-MRI.
The results showed that our proposed UMA-Net could learn from diverse scenarios and achieve state-of-the-art segmentation performance.


Our main contributions can be summarized as follows:

\begin{itemize}



\item We proposed UMA-Net, a novel medical image segmentation network that can effectively learn from diverse sources of annotations.  

\item We proposed AUEM and QAM to guide the learning of UMA-Net with uncertainty guidance from the pixel level and the image level, respectively.

\item We conducted extensive experiments on different datasets, covering 2D and 3D segmentation tasks, which showed that our proposed UMA-Net achieved state-of-the-art performance for diverse scenarios.
\end{itemize}


\section{Related Works}
\subsection{Medical Image Segmentation}

With the development of CNNs, many deep-learning architectures have been developed for medical segmentation tasks\cite{ronneberger2015u,9053405,2020UNet,huang2017densely,li2018h,cciccek20163d}. UNet \cite{ronneberger2015u}, and its variants have surprising performance and the widest application range among them. The distinguishing feature of the U-Net network is its U-shape structure and skip connection, which enables the resulting feature maps to capture not only low-level semantic information but also high-level semantic information, leading to superior segmentation results. Based on U-Net, Çiçek et al. \cite{cciccek20163d} proposed the 3D UNet, which uses 3D convolutions to replace the 2D convolutions in U-Net and can perform 3D medical image segmentation. Li et al. \cite{li2018h} proposed DenseU-Net by applying the Dense structure \cite{huang2017densely} to the U-Net network. Oktay et al. \cite{2020UNet} proposed the U-Net++ model, which introduces a built-in depth-variable collection of U-Nets that can improve segmentation performance for objects of different sizes. Based on this, Huang et al. \cite{9053405} proposed the UNet 3+ model, which adds full-scale skip connections on top of U-Net++, enabling the fusion of semantic information even across different levels of features and fully utilizing semantic information at different scales.

However, the aforementioned methods, as well as most learning-based approaches, require assigning a high-quality ground-truth annotation to each case in order to obtain a high-quality segmentation model and cannot directly learn from multi-source annotations.

\subsection{Learning from noisy annotations}
Learning from noisy labels is a challenging and critical task since the labels we get in many scenarios contain noise. Most existing research focuses on the classification task with noisy labels \cite{karimi2020deep}. For the image segmentation task, Zhu et al. \cite{zhu2019pick} developed a CNN that can differentiate between noisy and clean labels to facilitate learning from noisy labels in medical image segmentation. Similarly, Mirikharaji et al. \cite{mirikharaji2019learning} decreased the weights of pixels with loss gradient directions further from those of clean data. Nevertheless, both approaches necessitate a collection of clean labels for training. In \cite{min2019two}, a semi-supervised biomedical image segmentation method that employs an attention network was introduced to handle noisy labels. Meanwhile, D. Karimi et al.\cite{karimi2020deep} utilized dual CNNs with iterative label updates for fetal brain segmentation. In \cite{wang2020noise}, a novel network COPLE-Net with noise-robust Dice loss and self-ensembling was proposed for COVID-19 pneumonia lesion segmentation. TriNet \cite{zhang2020robust} designs a tri-network and uses integrated prediction from two networks for the third network training to alleviate the misleading problem. Z. Xu et al. \cite{xu2022anti} proposed a Mean-Teacher-assisted Confident Learning (MTCL) framework that learns from both low-quality sets and high-quality sets. In \cite{shi2021distilling}, it designs the image-level robust learning strategy according to the original image-level labels and pseudo labels and distills more effective information as a complement to pixel-level learning.

Nevertheless, the above works were based on single-source annotations and cannot utilize the information of multi-source annotations, such as commonality and difference information.

\subsection{Learning from Multi-source Annotations}

Due to the challenge of noises in single-source annotations, multi-source annotations are frequently used in clinical settings. The utilization of multi-source annotations for training segmentation networks has garnered researchers' interest in recent years. The majority vote involves averaging annotations, while STAPLE \cite{warfield2004simultaneous} employs an Expectation-Maximization (EM) framework to iteratively estimate the true segmentation and the performance levels of each rater. Both approaches merge multi-source annotations into a unique annotation and train the network accordingly. Label sampling \cite{jensen2019improving}, on the other hand, supervises the network by randomly selecting labels from the multi-rater labeling pool during each training iteration. Experimental results have demonstrated that label sampling better calibrates the model compared to using a single set of labels such as majority vote. Multiple-head/branch architecture models each individual rater by multiple decoders in a neural network, which also outperforms that of a unique combined ground truth. 
Recent works have made efforts to leverage both the commonality and difference information of multi-source annotations. MA-Net \cite{ji2021learning} used an expertise-aware inferring module to adapt to the different expertness of raters. Self-Calib \cite{wu2022learning} estimated the latent ground truth by recurrently running a diverging and converging model. MRIBNet \cite{zhang2023multi} introduced the information bottleneck to reduce redundant information and retain consistent information from multiple annotations. UMNet \cite{wang2022multi} used an uncertainty mining network to estimate the uncertainty maps of each annotation. These methods showed more abilities to leverage multi-annotation information than fusing tragedies. 

However, the above methods ignored the impact of low-quality samples. Some of them focus on learning reliable labels\cite{ji2021learning,wu2022learning}, but they do not identify low-quality samples, which led them to contaminate the network. Some methods simply discarded low-quality samples, which led to the loss of valuable information \cite{zhang2023multi}. 
In addition, none of these methods completely utilized multi-source annotation information, leaving room for optimization in subsequent work.

\begin{figure*}[t]
\centering
\includegraphics[width=1\textwidth]{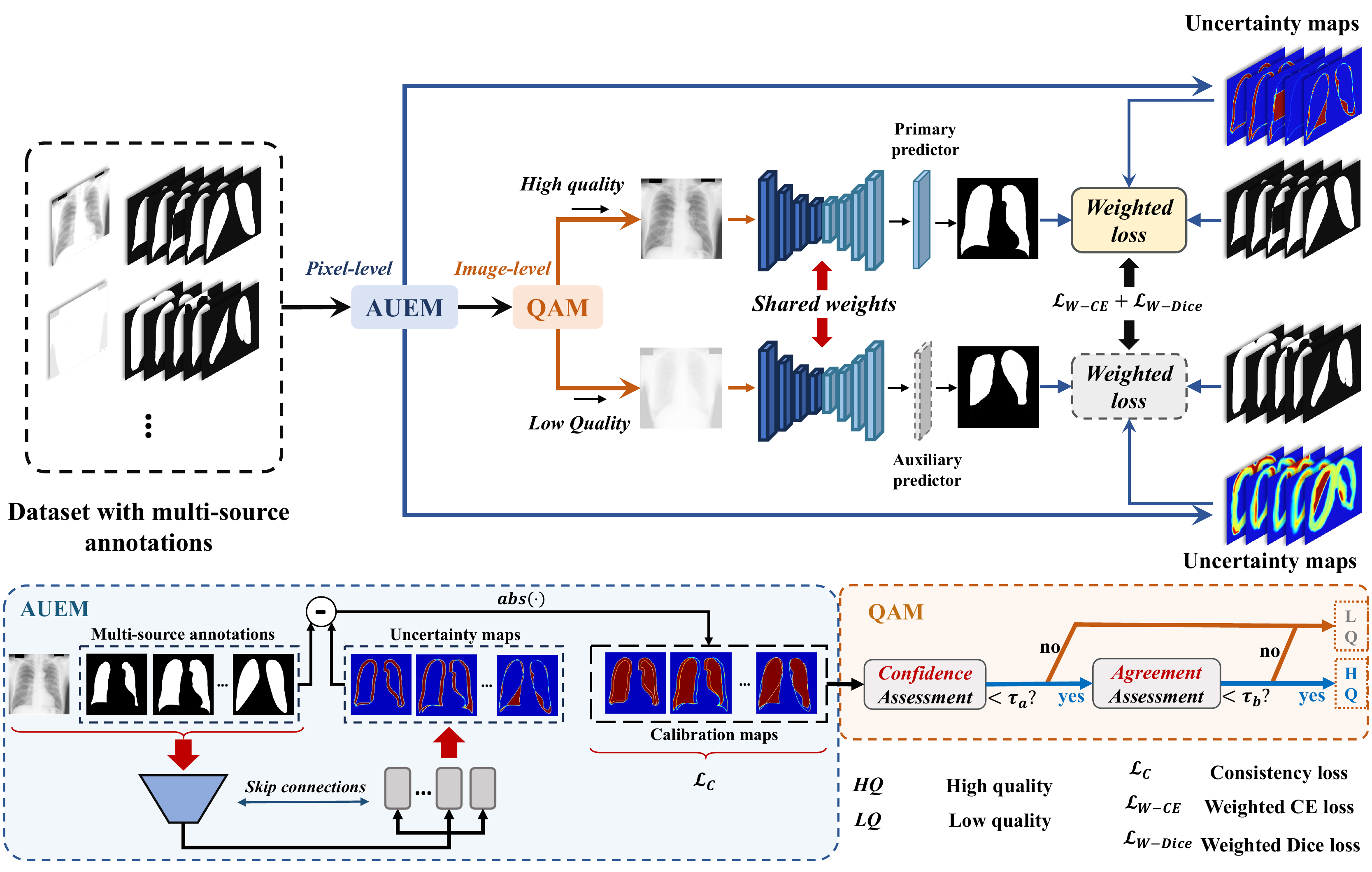}
\caption{The schematic illustration of our proposed UMA-Net, consisting of a segmentation backbone, e.g., UNet, with a primary predictor and an auxiliary predictor, an AUEM for pixel-level guidance, and a QAM for image-level guidance. Uncertainty maps are estimated by AUEM and then weigh the segmentation loss. QAM identifies the low-(high-)quality samples to be fed to the primary and auxiliary predictors, respectively.}
\label{fig:framework}
\end{figure*}

\section{Methodology}
\label{method}


In this section, we elaborate how we tackle the challenges of learning from multi-source annotations with the proposed Uncertainty-guided Multi-source Annotation Network (UMA-Net).
We first present the overall framework of UMA-Net, followed with technical details about the uncertainty guidance with the Annotation Uncertainty Estimation Module (AUEM) and the Quality Assessment Module (QAM).

\subsection{Overall Framework}

Our method is based on two observations. First, due to annotations from different sources containing different noise patterns, we are able to estimate the reliability of each pixel label by analyzing their consistencies and discrepancies. Second, directly learning from low-quality samples with a large noise ratio in annotations would lead to network degradation. 
These two above observations motivate us to develop an uncertainty-guided multi-source annotation network (UMA-Net) to learn the noise-robust segmentation network from multi-source annotations through pixel and image-level guidance.
As shown in Fig. \ref{fig:framework}, UMA-Net estimates pixel-wise uncertainty maps reflect the reliability of each segmentation annotations with an Annotation Uncertainty Estimation Module (AUEM) and weighs the segmentation loss by the uncertainty maps, thus guiding the network to learn from more reliable pixel labels. 
Meanwhile, UMA-Net assesses the image-level quality of the samples with a Quality Assessment Module (QAM) and uses an independent auxiliary predictor to learn from the low-quality samples, which avoids the error accumulation in the primary predictor to preserve its accuracy.
During testing, only the well-trained segmentation network with the primary predictor will be used, while other components can be discarded, which saves computational costs as well.

\subsection{Uncertainty-guided Learning from Multi-source Annotations}

\subsubsection{Annotation Uncertainty Estimation Module}
We estimated the uncertainty map of each annotation to reflect the reliability of each pixel-wise label. The uncertainty of single-source noisy annotation is difficult to estimate without reference. However, when multi-source annotations are available, due to the different noise patterns from different sources, it is possible to obtain the uncertainty of each pixel label by cross-referencing. 
Based on this observation, we designed the Annotation Uncertainty Estimation Module (AUEM) to estimate the pixel-wise uncertainty map for each annotation by analyzing the variation among multi-source annotations. The uncertainty maps then guide the segmentation network to learn reliable pixel labels by weighing the segmentation loss.

Specifically, AUEM consists of an encoder with multiple decoders, with each decoder serving as an independent uncertainty estimator for each annotation, as shown in Fig. \ref{fig:framework}. The input of the AUEM is the original image concatenated with all collected annotations. The output layers of the AUEM are the Sigmoid unit so that the estimated uncertainties are all within 0 and 1. The AUEM estimates pixel-wise uncertainty maps of annotations and then weighs the segmentation loss of the UNet to guide the UNet only learns from reliable labels. 
According to \cite{wang2022multi}, by jointly training the AUEM and the segmentation network with weighted loss, AUEM could spontaneously analyze the discrepancy among annotations and assign unreliable pixel labels to large values and reliable labels to small values. Formally, we train the model with a dataset $\mathcal{D}=\{x_k, y_k^1, y_k^2, ..., y_k^M\}_{k=1}^{\Omega}$ that contains $M$ annotations per sample, where $y_k^m$ means the $m^{th}$ annotation of the sample $x_k$. Assuming $p_i$ is the $i^{th}$ pixel of the prediction of the UNet, and $\sigma_i^m$ is the estimated uncertainty value of $i^{th}$ pixel on $m^{th}$ annotation. Following\cite{wang2022multi}, the weighted cross entropy (W-CE) loss $\mathcal{L}_{\rm W\text{-}CE}$ can be derived as follows:

\begin{equation}\label{ce_loss}
    \mathcal{L}_{\rm W\text{-}CE} = \frac{1}{N\times M}\sum_{m=1}^M \sum_{i=1}^{N} (\frac{1}{exp(\sigma_i^m)}\mathcal{L}_{\rm CE}(p_i, y_i^m)+\frac{1}{2}\sigma_i^m)
\end{equation}

\noindent where $N$ is the total number of pixels/voxels per image, and the coefficient $ \frac{1}{exp(\sigma_i^m)} $ decreases the contribution of unreliable pixels with large uncertainty values to the final loss. The term $\frac{1}{2}\sigma_i^m$ is a regularization term to prevent uncertainty values from being too large.

When segmenting small objects, using CE loss can make CNNs biased toward learning the background. A common approach is to combine it with the Dice loss. Therefore, we redesign the Dice loss to become an uncertainty-weighted Dice loss. The original Dice loss is represented as follows:

\begin{equation}\label{dice_loss}
    \mathcal{L}_{Dice}=1-\frac{2\sum_i^{N}p_iy_i}{\sum_i^{N} p_i^2+\sum_i^{N} {y_i}^2}=\frac{\sum_i^{N}(p_i-y_i)^2}{\sum_i^{N} p_i^2+\sum_i^{N} {y_i}^2}
\end{equation}

By applying pixel-wise weighing on Dice loss using the uncertainty map, we obtained the expression for the weighted Dice loss as follows:

\begin{multline}
    \mathcal{L}_{\rm W\text{-}Dice}=\frac{1}{ M}\sum_{m=1}^M\sum_{i=1}^{N} \\ (\frac{1}{exp(\sigma_i^m)}\frac{(p_i-y_i^m)^2}{\sum_i^{N} (p_i)^2+\sum_i^{N} ({y_i^m})^2}+\frac{1}{2}\frac{\sigma_i^m}{N})
\end{multline}

\subsubsection{Achieve an Agreement of AUEM}
To further improve AUEM's internal cross-reference ability and facilitate uncertainty estimators of the AUEM to reach an agreement on the latent ground truth, we proposed the concept of the calibration map. The calibration map is the coarse estimation of latent ground truth obtained through the noisy annotation and its corresponding uncertainty map. The calculation of the calibration map is based on the assumption that a noisy label with lower uncertainty has a higher probability of being the correct one, and vice versa\cite{wu2021uncertainty}. The calculation formula of calibration map $s$ is as follows:

\begin{equation}
s_i^m= |y_i^m-\sigma_i^m| \quad and \quad s_i^*=\frac{1}{M}\sum^M_{m=1}s_i^m
\end{equation}

The calibration map $s^m$ represents the coarse estimated latent ground truth implied by the $m_{th}$ estimator, and $s^*$ is the average calibration map. However, the calibration maps from different uncertainty estimators may vary a lot. We introduce an image-level consistency regularization by minimizing the L2 loss among these calibration maps. The consistency loss forces the AUEM to improve its internal cross-reference ability to analyze the variation of the annotations and facilitates the estimators of AUEM to reach an agreement on the latent ground truth:

\begin{equation}
\mathcal{L}_{\rm C}=\frac{1}{M\times N}\sum_{m=1}^M \sum_{i=1}^N \Vert s_i^m-s_i^* \Vert _2^2
\end{equation}

\subsection{Decoupling of Low-quality Samples}
\label{sec:Quality Assessment Module}

We define samples with a large noise ratio among multi-source annotations as low-quality samples. 
In our method, learning from these low-quality samples could harm the performance of the network. 
The reason is that a large number of middle values will appear in the uncertainty maps that AUEM estimates from these low-quality samples. These uncertainties with middle values cannot guide the network to maximize the learning of correct pixel labels and minimize the learning of wrong labels. 
In addition, even with the constraint of consistent loss $L_C$, there will still be a lot of conflicting information in their uncertainty maps, which can be reflected in the excessive inconsistency among calibration maps. 
To alleviate this problem, we first design a quality assessment module (QAM) to identify (high-)low-quality. High-quality samples will be directly learned by the UNet with its primary predictor. 
We further add an auxiliary predictor of the segmentation network to exploit the remaining low-quality samples without sacrificing the accuracy of the primary predictor. 
The following sections will provide introductions to the Quality Assessment Module (QAM) and the segmentation network with an auxiliary predictor.

\subsubsection{Quality Assessment Module}

The Quality Assessment Module (QAM) identifies (high-) low-quality samples through the calibration maps. 
Under two situations, we consider the input sample as a low-quality sample. 
The first is when a substantial number of pixel labels in the calibration maps are the middle values instead of being closer to 0 or 1. The other is when significant conflicts arise among the calibration maps of the sample. We quantified the two situations through Confidence Assessment and Agreement Assessment, respectively.

\textbf{Confidence Assessment} quantifies the confidence score by calculating the entropy of the mean calibration map. The quantification formula is as follows:

\begin{equation}
u_a=\frac{ \sum_{i=1}^{N} (-s_{i=1}^*logs_{i=1}^*)}{ \sum_{i=1}^{N} s_i^*}
\end{equation}

According to the formula, it can be observed that when there is a large number of estimated values around 0.5 in the calibration map, the confidence score will be high. On the other hand, samples with lower scores can be considered to have higher quality.

\textbf{Agreement Assessment} obtains the agreement score by calculating the variance among calibration maps. The calculation formula is as follows:

\begin{equation}
u_b=\frac{4M \cdot \sum_{i=1}^{N} {\rm Var}(s_i^1, s_i^2, ...,s_i^M) }{\sum_{m=1}^M \sum_{i=1}^{N} s_i^m}
\end{equation}

\noindent where coefficient $4M$ is used to map the resulting score to a range between 0 and 1. According to the formula, the agreement score will be high if there are too many contradictions among calibration maps. 

We determine a sample as a high-quality sample by setting two thresholds $\tau_a$ and $\tau_b$. The samples that correspond to $u_a<\tau_a $ and $ u_b<\tau_b$ are determined to be high-quality samples, and others belong to low-quality samples. 
The segmentation network with its primary predictor will learn from these identified high-quality samples.

\subsubsection{Auxiliary predictor for preventing error accumulation}

Only learning from high-quality samples is insufficient to achieve an optimal solution. 
Leveraging low-quality data could significantly enhance the effectiveness of data utilization. Previous work such as \cite{ji2021learning,wu2022learning} did not design to identify low-quality samples, leaving them to contaminate the network. On the other hand, the research \cite{zhu2019pick,zhang2023multi} directly discarded low-quality samples, leading to a significant loss of valuable data.

To tackle this problem, we add an additional auxiliary predictor of the network independent from the primary predictor. The auxiliary predictor and the primary predictor are fed with features from the same backbone network. During the training phase, the QAM is first used to identify the quality of each sample. 
Then, the sample is fed to the primary predictor or the auxiliary predictor depending on whether it is identified of high quality or low quality, respectively.
In this way, the primary predictor is prevented from error accumulation caused by low-quality data, while the backbone could still learn representation knowledge from those data with the auxiliary predictor.
Combining QAM and the auxiliary predictor, we manage to guide the network learning at the image level.

\subsection{Overall Training Procedure}
The final loss of the UMA-Net is the combination of the weighted cross entropy loss, the weighted Dice loss, and the consistency loss:
\begin{equation}
\label{overall_loss}
\mathcal{L}= \mathcal{L}_{\rm W\text{-}CE}+\lambda \mathcal{L}_{\rm W\text{-}Dice}+\alpha \mathcal{L}_{\rm C},
\end{equation}

where $\lambda$ and $\alpha$ are hyper-parameters.
Note that the two heads of UMA-Net are both trained with Eq. \ref{overall_loss}, and QAM determines the samples used for each head.
The whole objective can be optimized end-to-end.

\section{Experiments}\label{experiments}

Extensive experiments were conducted on three datasets: 1. the JSRT \cite{shiraishi2000development} dataset for lung segmentation from chest X-rays; 2. the Duke-Breast-Cancer-MRI \cite{saha2018machine} dataset for breast cancer segmentation from DCE-MRI; and 3. the RIGA \cite{almazroa2017agreement} dataset for optic disc and cup segmentation from fundus images.


All the networks involved in our experiments were implemented using Pytorch on a server with an NVIDIA Tesla V100 GPU (32 GB). The backbone of the segmentation network is the 3D UNet for Duke-Breast-Cancer-MRI and the 2D UNet for JSRT and RIGA datasets. 
For training, we use Adam as the optimizer with an initial learning rate of 0.001 and a learning decay rate of 0.96 per epoch. $\alpha$ is set to be a sigmoid-shape monotonically function of the training steps with a maximum of 1. $\lambda$ is set to 1. Thresholds $\tau_a$ and $\tau_b$ for the breast cancer MRI segmentation are set to 0.15 and 0.1, and for the other 3 datasets are set to 0.2 and 0.2. 
For testing, we only take the prediction of the segmentation network with the primary predictor.

\begin{figure}[t]
  \centering
  \includegraphics[width=\linewidth]{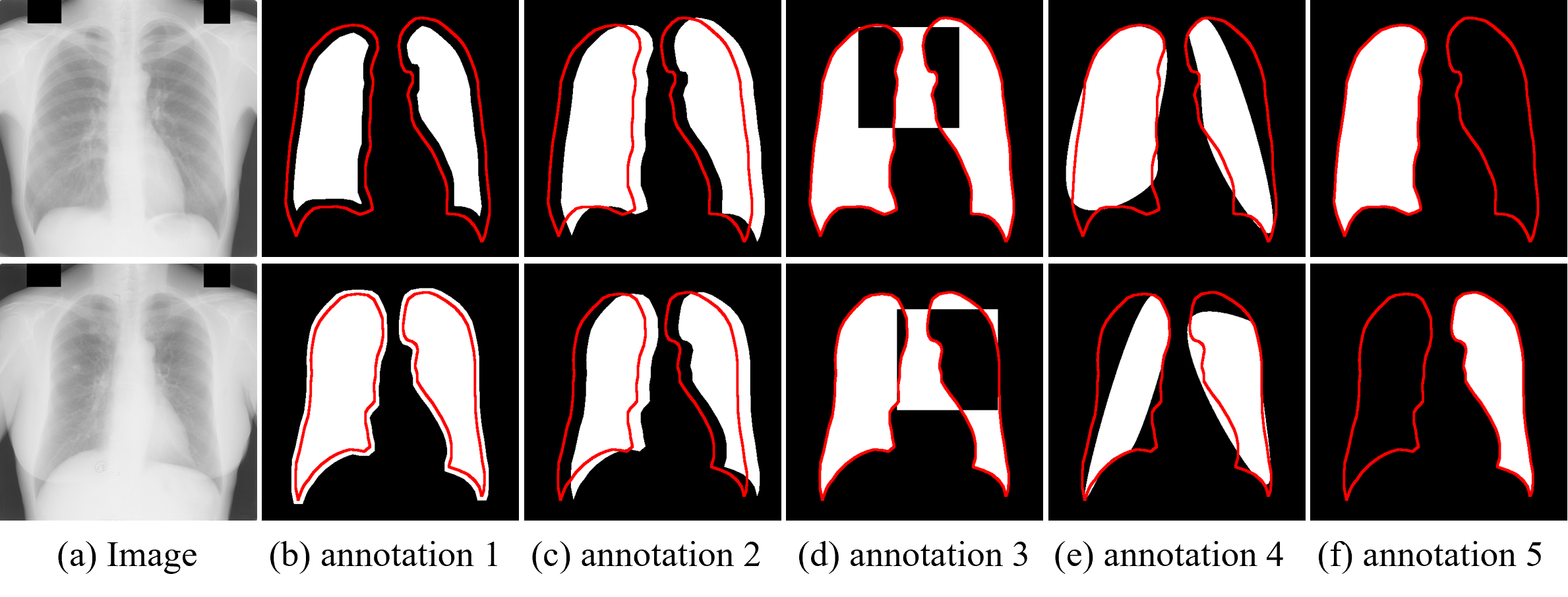}
  \caption{Two chest X-ray samples, each has five different annotations with different types of noise. Red lines delineates the ground truth.}
  \label{jsrt noisy annotations}
\end{figure}

\begin{table}[]
\caption{Segmentation performance by leveraging different types of annotations for lung segmentation from chest X-rays.}\label{Lung comparison result}
\resizebox{0.49\textwidth}{!}{
\centering
\begin{tabular}{ccccc|cc}
\hline
\toprule[2pt]
\multicolumn{5}{c|}{Annotation Set}                            & \multicolumn{2}{c}{metrics}            \\ \hline
Ann. 1     & Ann. 2     & Ann. 3     & Ann. 4     & Ann. 5     & \textit{Dice(\%)$\uparrow$} & \textit{Jaccard(\%)$\uparrow$} \\ \hline
\checkmark & \checkmark &            &            &            & 80.03            & 66.71               \\
\checkmark & \checkmark & \checkmark &            &            & 86.29            & 75.89               \\
\checkmark & \checkmark & \checkmark & \checkmark &            & 86.91            & 76.85               \\
\checkmark & \checkmark & \checkmark & \checkmark & \checkmark & 88.75            & 79.78               \\
\checkmark &            & \checkmark &            &            & 84.99            & 73.90               \\
\checkmark &            &            & \checkmark &            & 79.51            & 65.99               \\
\checkmark &            &            &            & \checkmark & 82.00            & 69.49               \\ 
\bottomrule[2pt]
\end{tabular}
}
\end{table}

\begin{table*}[]\caption{Comparison of our proposed method and other state-of-the-art approaches on the LN-JSRT and HN-JSRT datasets. The average and standard deviation of three runs are reported. The best performance is highlighted in \textbf{bold}. MV means majority vote. $^*$ means the network was trained by the undistorted annotations.}\label{Lung result}
\centering
\resizebox{\textwidth}{!}{
\begin{tabular}{c|cccccccc}
\hline
\toprule[2pt]
\multirow{3}{*}{\textbf{Method}} & \multicolumn{8}{c}{\textbf{Lung segmentation}}                                                                                                                                                     \\ \cline{2-9} 
                                 & \multicolumn{4}{c|}{\textbf{LN-JSRT}}                                                                    & \multicolumn{4}{c}{\textbf{HN-JSRT}}                                               \\ \cline{2-9} 
                                 & \textit{Dice(\%)$\uparrow$} & \textit{Jaccard(\%)$\uparrow$} & \textit{ASD(pixel)$\downarrow$} & \multicolumn{1}{c|}{\textit{95HD(pixel)}$\downarrow$} & \textit{Dice(\%)$\uparrow$} & \textit{Jaccard(\%)$\uparrow$} & \textit{ASD(pixel)$\downarrow$} & \textit{95HD(pixel)$\downarrow$} \\ \hline
\textbf{UNet$^*$}                                & 96.23 $\pm$ 0.4             & 92.73 $\pm$ 0.7                & 11.79 $\pm$ 1.5               & \multicolumn{1}{c|}{30.74 $\pm$ 4.0}                & 96.63 $\pm$ 0.4             & 92.73 $\pm$ 0.7                & 11.79 $\pm$ 1.5               & 30.74 $\pm$ 4.0                \\
\textbf{UNet (MV)\cite{ronneberger2015u}}                    & 82.23 $\pm$ 2.3             & 69.82 $\pm$ 2.4                & 17.46 $\pm$ 4.7               & \multicolumn{1}{c|}{52.93 $\pm$ 7.3}                & 75.28 $\pm$ 4.1             & 60.36 $\pm$ 2.9                & 21.66 $\pm$ 5.2               & 67.92 $\pm$ 9.4                \\
\textbf{nnUNet (MV)\cite{isensee2021nnu}}                    & 82.67 $\pm$ 2.2             & 70.46 $\pm$ 3.2                & 12.93 $\pm$ 3.2               & \multicolumn{1}{c|}{49.50 $\pm$ 7.2}                & 76.85 $\pm$ 3.9             & 62.40 $\pm$ 3.7                & 17.73 $\pm$ 3.7               & 69.13 $\pm$ 9.3                \\
\textbf{COPLE-Net (MV)\cite{wang2020noise}}                  & 87.19 $\pm$ 1.7             & 77.29 $\pm$ 2.6                & 13.37 $\pm$ 3.0               & \multicolumn{1}{c|}{44.66 $\pm$ 6.1}                & 81.58 $\pm$ 2.3             & 68.89 $\pm$ 3.1                & 16.66 $\pm$ 3.6               & 58.31 $\pm$ 8.1                \\
\textbf{PINT (MV)\cite{shi2021distilling}}                   & 87.66 $\pm$ 1.8             & 78.03 $\pm$ 2.8                & 12.19 $\pm$ 2.5               & \multicolumn{1}{c|}{46.90 $\pm$ 6.2}                & 82.96 $\pm$ 2.2             & 70.88 $\pm$ 3.7                & 15.59 $\pm$ 2.9               & 56.37 $\pm$ 8.3                \\
\textbf{MR-Net\cite{ji2021learning}}                    & 89.10 $\pm$ 1.5             & 80.34 $\pm$ 1.9                & 12.80 $\pm$ 2.5               & \multicolumn{1}{c|}{41.54 $\pm$ 5.5}                & 85.43 $\pm$ 1.9             & 74.56 $\pm$ 2.1                & 14.51 $\pm$ 2.9               & 51.98 $\pm$ 7.4                \\
\textbf{Self-Calib\cite{wu2022learning}}                & 91.50 $\pm$ 1.1             & 84.33 $\pm$ 1.5                & 12.09 $\pm$ 2.4               & \multicolumn{1}{c|}{43.29 $\pm$ 5.8}                & 85.39 $\pm$ 1.5             & 74.50 $\pm$ 2.0                & 13.88 $\pm$ 2.8               & 51.33 $\pm$ 7.8                \\ \hline
\textbf{UMA-Net (Ours)}                              & \textbf{92.97 $\pm$ 0.9}    & \textbf{86.86 $\pm$ 1.2}       & \textbf{9.39 $\pm$ 1.7}       & \multicolumn{1}{c|}{\textbf{36.98 $\pm$ 4.9}}       & \textbf{88.75 $\pm$ 1.1}    & \textbf{79.78 $\pm$ 1.5}       & \textbf{11.91 $\pm$ 2.3}      & \textbf{42.04 $\pm$ 6.6}       \\
\bottomrule[2pt]
\end{tabular}
}
\end{table*}

\subsection{Lung Segmentation with Simulated Annotations}

\subsubsection{Datasets}
We prepare two simulated datasets, LN-JSRT and HN-JSRT, derived from the original JSRT \cite{shiraishi2000development} dataset for lung segmentation. Both LN-JSRT and HN-JSRT have all 247 x-ray scans from JSRT that are split into 165 training scans, 12 validation scans, and 70 testing scans. To simulate annotations from different sources, the training scans are paired with five noisy annotations derived from expert-rated clean annotation by adding various types of noise to the clean annotation, as shown in Fig. 2. LN-JSRT has a lower noise level with the average Dice score of its noisy annotations is 0.8, while HN-JSRT has a higher noise level, and the average Dice score of the annotations is 0.7. 

For annotations from different sources, we add five types of noise:
1. \textbf{Random erosion or dilation.} Apply erosion or dilation to the clean mask with a 50\% probability for each operation.
2. \textbf{Constant Shift.} The masks are constantly shifted to the right for several pixels.
3. \textbf{Square reverse.} Reverse the labels in a randomly picked square.
4. \textbf{Polygonization.} Choose several points on the boundary of the clear mask and reconnect them by the cubic spline interpolation function. 
5.\textbf{Random erase.} Randomly erase one lung mask in the annotation with a constant probability.

\begin{figure*}[t]
  \centering
  \includegraphics[width=\linewidth]{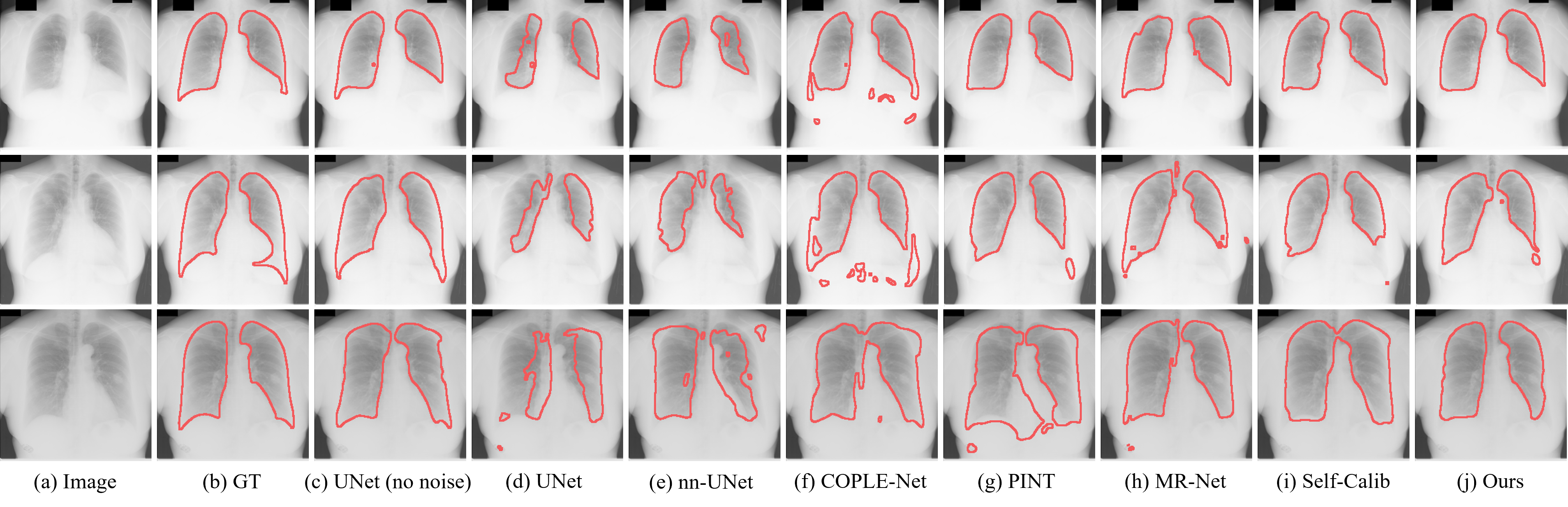}
  \caption{Qualitative comparison of our proposed UMA-Net and other SOTA approaches on HN-JSRT.}
  \label{fig:seg_nodule}
\end{figure*}

\subsubsection{Comparison with the State-of-the-arts}
We compared our proposed method with other state-of-the-art segmentation approaches, including UNet\cite{ronneberger2015u}, nnUNet\cite{isensee2021nnu}, COPLE-Net\cite{wang2020noise}, PINT\cite{shi2021distilling}, MR-Net\cite{ji2021learning}, and Self-Calib\cite{wu2022learning}. 
Among them, UNet and nnUNet are general single-annotation frameworks, COPLE-Net and PINT are noise-robust single-annotation frameworks, and MR-Net and Self-Calib are the current SOTA multi-source annotation methods. 
For the single-annotation frameworks (UNet, nnUNet, COPLE-Net, and PINT), we utilized the majority vote strategy on five noisy annotations to obtain final annotation for training. 
We evaluated the segmentation performance using four metrics: Dice similarity coefficient \textbf{(DSC)}, Jaccard similarity coefficient \textbf{(JSC)}, Average Surface Distance \textbf{(ASD)}, and 95\% Hausdorff Distance \textbf{(95HD)}. 
We conducted experiments on our proposed method using both the low-noise dataset LN-JSRT and the high-noise dataset HN-JSRT. The results in Table \ref{Lung result} show that our method outperforms other state-of-the-art methods at both noise levels. 
For the low noise dataset LN-JSRT, our method achieves a Dice score of 92.97\%, which is 10.74\% higher than directly training UNet with the majority vote strategy. Using a noise-robust framework is better than UNet but still lower than our method by 5.31\%. 
Our method also defeats the other two SOTA multi-source annotation methods with at least a 1.47\% lead. 
For the high-noise dataset HN-JSRT, our lead continues to increase. Our method achieves a Dice score of 88.75\% and outperforms PINT and Self-Calib by 5.79\% and  3.36\%, respectively. 

Qualitative results are illustrated in Fig. \ref{fig:seg_nodule}. The results show that our method achieves better segmentation performance than noise-robust single-label frameworks and other multi-source annotation methods at both noise levels.

\subsubsection{Influence of Annotation Quantity and Noise Type}

To further investigate the influence of annotation quantity and noise type on the performance of our proposed method, we conducted comparative experiments by training the model with different annotation combinations in HN-JSRT. The results are shown in Table \ref{Lung comparison result}, which indicates that annotation quantity and noise type have a decisive impact on the model's performance. First, as the number of annotations increases from 2 to 5, the model's performance gradually improves from 80.03\% dice score to 88.75\% dice score. Moreover, when the annotations per image are kept at two, different annotation combinations have different effects on the model's performance, such as the combination of \textbf{Random Erosion/Dilation} and \textbf{Square Area Inversion} is better than the combination of \textbf{Random Erosion/Dilation} and \textbf{Polygonization}, which suggests that noise type of annotations greatly influences the performance of our model.

\subsection{RIGA Dataset}

\subsubsection{Datasets} 
Inter-observer variation and intra-observer variation are significant in optic nerve head segmentation. Several raters are usually needed to alleviate these problems. 
In this work, we use the publicly released benchmarks RIGA \cite{almazroa2017agreement}  for optic disc and cup (OD/OC) segmentation, which contains a total of 750 fundus images, including 460 images from MESSIDOR, 195 images from BinRushed, and 95 images from Magrabia. Six glaucoma experts independently labeled the optic cup and disc contour masks manually. A senior expert references their masks and arbitrates the final ground truth. All images are resized to 256×256 as input of the network. Following previous works \cite{ji2021learning,wu2022learning}, We select 655 samples from BinRushed and MESSIDOR for training, and 95 samples from Magrabia for testing.

\subsubsection{Comparison with the State-of-the-arts}

A quantitative comparison on RIGA dataset can be found in Table \ref{RIGA result}. 
Due to the individual preferences of different raters, some rater-related biased noise is ingrained in each annotation. Annotations from different raters can vary widely, especially for optic cup segmentation because of its blurred boundaries. 
We compare our proposed UMA-Net with MR-Net \cite{ji2021learning}, Self-Calib \cite{wu2022learning}, AGNet \cite{zhang2019attention} and BEAL \cite{wang2019boundary}. 
Specifically, UMA-Net surpasses all the compared approaches with more than 0.13\% dice score in optic disc segmentation and more than 0.55\% dice score in optic cup segmentation, achieving a new state-of-the-art performance on this dataset. Even when the boundary of the optic cup is highly uncertain, our method can give results closer to ground truth. 
The results further demonstrate UMA-Net's effectiveness in learning from multi-source annotations that contain individual bias and large inter-rater variations.

\begin{figure}[t]
  \centering
  \includegraphics[width=\linewidth]{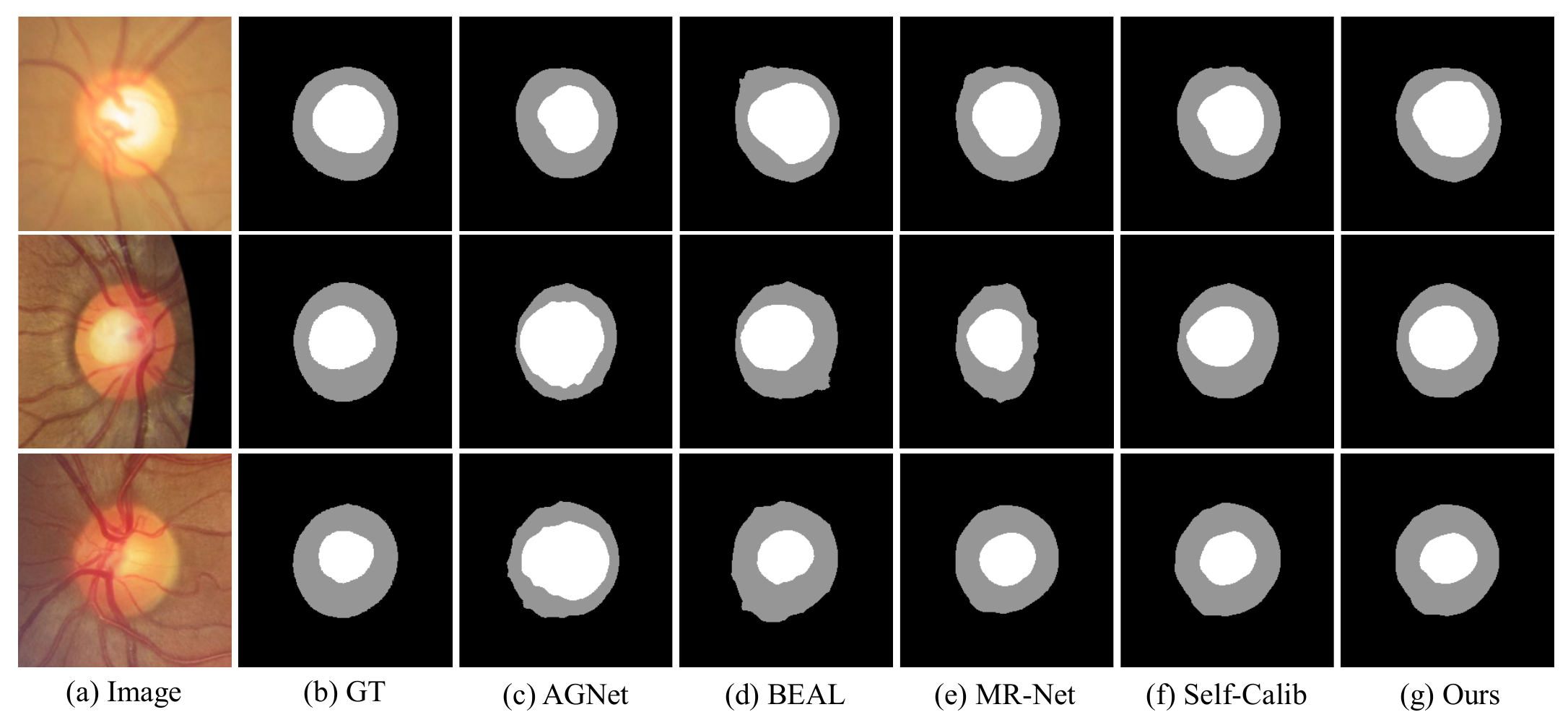}
  \caption{Qualitative comparison of optic disc/cup segmentation among other SOTA approaches and our method.}\label{RIGA Image}
\end{figure}

The qualitative comparison is illustrated in Fig. \ref{RIGA Image}. The visualized segmentation results show that UMA-Net could detect and delineate both optic disc and cup more accurately than previous optic disc/cup segmentation methods.

\subsection{Breast DCE-MRI Cancer Segmentation}

\subsubsection{Datasets}
Breast cancer is a severe global health challenge \cite{luo2023deep}. Breast cancer segmentation plays an important role in treatment planning, whereas accurate annotations are hard to obtain \cite{zhou2019weakly,luo2019deep}. Hence, we proposed to segment breast cancers from DCE-MRI using multi-source annotations obtained by multiple automated algorithms. 
In this work, we use the publicly available Duke-Breast-Cancer-MRI dataset \cite{saha2018machine} that provides 922 3D dynamic contrast-enhanced magnetic resonance images (DCE-MRI) with malignant tumors. We select 175 cases and invite an experienced radiologist to annotate the tumor mask as ground truth. We segment the breast area by the method in \cite{wei2018three} to eliminate the impact of the chest. The pre-contrast image and the first two post-contrast images are resampled to 0.7mm×0.7mm×1.4mm voxel spacing, cropped to 320×224×160, and then concatenated as input. The training set consists of 90 cases, the validation set has 10 cases, and the testing set has 75 cases. 

Four different region-growing-based algorithms \cite{militello2021unsupervised} are implemented as annotation sources to generate four pseudo annotations per sample. Each algorithm requires seven steps: breast region segmentation, subtraction image calculation, blood vessel estimation, blood vessel suppression, tumor enhancement map estimation, growth seed setting, and region growing. The first five steps are the same, but the last two steps have some differences, which cause their characteristic differences, i.e., Algorithm 1 has the great ability to track boundaries but is sensitive to noise, while Algorithm 2 is more robust to noise but unable to track boundary details. Due to the different characteristics of the algorithms, annotations generated by different algorithms have their distinctive systematic noise pattern.
It is worth noting that the annotations rated by experts are only used to test the model performance and do not participate in the training stage.

\begin{table}[]
\centering
\caption{Performance comparison on the RIGA dataset. The best performance is highlighted in \textbf{bold}. MV means majority vote.}\label{RIGA result}
\resizebox{0.43\textwidth}{!}{
\begin{tabular}{c|cccc}
\hline
\toprule[2pt]
\multirow{2}{*}{\textbf{Method}} & \multicolumn{2}{c|}{\textbf{disc}}                        & \multicolumn{2}{c}{\textbf{cup}}     \\ \cline{2-5} 
                                 & \textit{Dice(\%)$\uparrow$} & \multicolumn{1}{c|}{\textit{IoU(\%)}$\uparrow$} & \textit{Dice(\%)$\uparrow$} & \textit{IoU(\%)$\uparrow$} \\ \hline
\textbf{AGNet (MV)\cite{zhang2019attention}}                   & 96.31             & \multicolumn{1}{c|}{92.93}            & 72.05             & 59.44            \\
\textbf{BEAL (MV)\cite{wang2019boundary}}                    & 97.08             & \multicolumn{1}{c|}{94.38}            & 85.97             & 77.18            \\
\textbf{MR-Net\cite{ji2021learning}}                  & 97.55             & \multicolumn{1}{c|}{95.24}            & 87.20             & 78.62            \\
\textbf{Self-Calib\cite{wu2022learning}}              & 97.82             & \multicolumn{1}{c|}{95.73}            & 90.15             & 82.07            \\ \hline
\textbf{UMA-Net (Ours)}          & \textbf{97.95}    & \multicolumn{1}{c|}{\textbf{95.98}}   & \textbf{90.70}    & \textbf{83.02}   \\
\bottomrule[2pt]
\end{tabular}
}
\end{table}

\subsubsection{Comparison with the State-of-the-arts}
Quantitative comparison on Duke-Breast-Cancer-MRI dataset between UMA-Net and six baselines are reported in Table \ref{Breast result}. The UNet and two noise-robust frameworks COPLE-Net \cite{wang2020noise} and PINT \cite{shi2021distilling} are trained with majority vote annotations. Multi-source annotation methods MR-Net \cite{ji2021learning} and Self-Calib \cite{wu2022learning} are directly trained with four pseudo annotations generated by four algorithms. The training result of the UNet supervised by ground truth is shown in the first row as an upper bound. As observed, UMA-Net outperforms other methods with at least 1.42\% in Dice and 1.96\% in Jaccard, 0.9 in ASD, and 4.1 in 95HD, which demonstrates its effectiveness. It is noteworthy that since all pseudo annotations are generated automatically, the whole pipeline is in an unsupervised manner. Even though, our method can approach the level of UNet with supervised pipeline, with only 2.75\% lower in Dice. To take full advantage of unsupervised learning, we further add the remaining unlabeled 747 cases in Duke-Breast-Cancer-MRI dataset to the training dataset and report the results in the last row of Table \ref{Breast result}. The result indicates that our proposed UMA-Net is an effective approach to learning from multiple sources of algorithm-generated pseudo labels for unsupervised segmentation and has the ability to scale up to larger datasets.

\begin{table}[]\caption{Performance comparison on the Duke-Breast-Cancer-MRI dataset. The average and standard deviation of three runs are reported. The best performance is highlighted in \textbf{bold}.\\ MV means majority vote. $^*$ means the network was trained by expert annotations. $^\dagger$ means the network was trained with the extended dataset.} \label{Breast result}
\centering
\resizebox{0.5\textwidth}{!}{
\begin{tabular}{c|cccccccc}
\hline
\toprule[2pt]
\multirow{1}{*}{\textbf{Method}}                           & $Dice(\%)\uparrow$                         & $~Jaccard(\%)\uparrow$                                & $ASD\downarrow$                  & $95HD\downarrow$                 \\ \hline
\textbf{UNet$^*$}                                      & 83.81 $\pm$ 0.11                           & ~~72.44  $\pm$ 0.12                          & 1.7  $\pm$ 0.2                          & 6.2 $\pm$ 0.7                           \\
\textbf{UNet (MV)\cite{ronneberger2015u}}                                              & 77.34  $\pm$ 0.40                          & ~~63.34  $\pm$ 0.45                          & 5.1  $\pm$ 0.7                          & 17.7 $\pm$ 2.7                          \\
\textbf{COPLE-Net (MV)\cite{wang2020noise}}   & 78.42 $\pm$ 0.31                           & ~~64.74  $\pm$ 0.35                          & 4.7 $\pm$ 0.6                           & 15.2 $\pm$ 2.3                          \\
\textbf{PINT (MV) \cite{shi2021distilling}}       & 78.76 $\pm$ 0.42                           & ~~65.23  $\pm$ 0.66                          & 4.5 $\pm$ 0.9                           & 14.9 $\pm$ 3.2                          \\
\textbf{MR-Net \cite{ji2021learning}}    & 79.25 $\pm$ 0.17                           & ~~65.92 $\pm$ 0.20                           & 4.4 $\pm$ 0.7                           & 14.5 $\pm$ 2.1                          \\
\textbf{Self-Calib \cite{wu2022learning}} & 79.64 $\pm$ 0.17                           & ~~66.56 $\pm$ 0.21                           & 3.5 $\pm$ 0.6                           & 13.5 $\pm$ 1.8                          \\ \hline
\textbf{UMA-Net (Ours)}                                    & \textbf{81.06 $\pm$ 0.14} & ~~\textbf{68.52 $\pm$ 0.15} & \textbf{2.6 $\pm$ 0.5} & \textbf{9.4 $\pm$ 1.6} \\
\textbf{\makecell{UMA-Net$^\dagger$ (Ours) \\ }}                               & \textbf{82.44 $\pm$ 0.10} & ~~\textbf{70.37 $\pm$ 0.12} & \textbf{2.2 $\pm$ 0.5} & \textbf{8.1 $\pm$ 1.4} \\
\bottomrule[2pt]
\end{tabular}
}
\end{table}

\begin{figure}[t]
  \centering
  \includegraphics[width=1\linewidth]{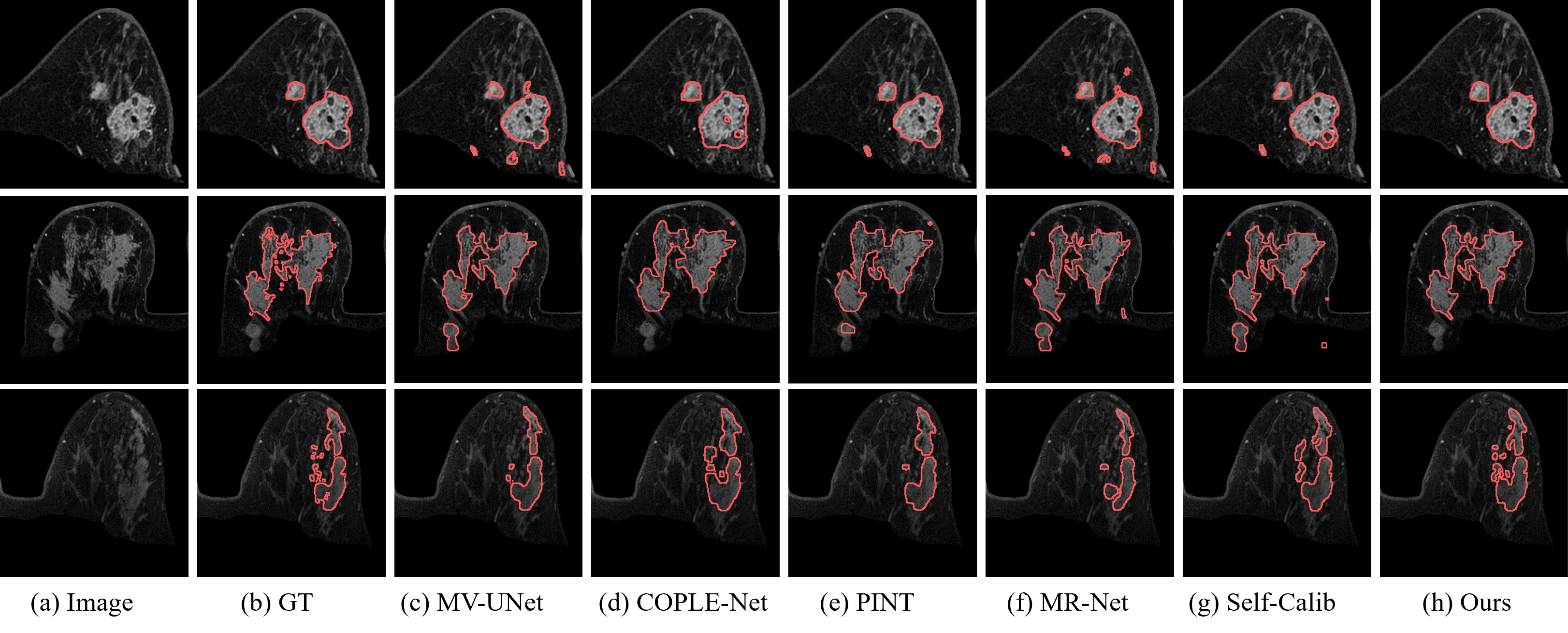}
  \caption{Qualitative comparison on breast cancer segmentation among other SOTA approaches and our method. }
  \label{Breast Image}
\end{figure}

Qualitative results are illustrated in Fig. \ref{Breast Image}. It can be seen that our proposed method is more accurate for ambiguous boundary segmentation. Meanwhile, our method can greatly reduce the false positives caused by blood vessels and skin highlight regions.

\begin{table*}[]
\centering
\caption{Ablation study on the components of our proposed UMA-Net. BL is the baseline by training UNet with majority vote. CL represents the consistency loss $L_c$. AP means the auxiliary predictor.}\label{Ablation result}
\resizebox{0.9\textwidth}{!}{
\begin{tabular}{ccccc|cc|cc|cc}
\hline
\toprule[2pt]
\multicolumn{5}{c|}{Components}                   & \multicolumn{2}{c|}{Lung}                & \multicolumn{2}{c|}{Breast cancer}       & \multicolumn{2}{c}{optic cup/disc}   \\ \hline
BL         & AUEM       & CL         & QAM        & AP        & \textit{Dice(\%)$\uparrow$} & \textit{Jaccard(\%)$\uparrow$} & \textit{Dice(\%)$\uparrow$} & \textit{Jaccard(\%)$\uparrow$} & \textit{Dice(\%)$\uparrow$} & \textit{IoU(\%)$\uparrow$} \\ \hline
\checkmark &            &            &            &            & 75.28             & 60.36                & 77.34             & 63.34                & 95.17/70.33       & 90.78/54.24      \\
\checkmark & \checkmark &            &            &            & 82.12             & 69.67                & 78.78             & 64.98                & 96.16/80.47       & 92.60/67.32      \\
\checkmark & \checkmark & \checkmark &            &            & 85.29             & 74.35                & 80.05             & 66.73                & 97.03/85.25       & 94.23/74.29      \\
\checkmark & \checkmark & \checkmark & \checkmark &            & 86.71             & 76.53                & 80.53             & 67.36                & 97.69/85.10       & 95.48/74.06      \\
\checkmark & \checkmark & \checkmark & \checkmark & \checkmark & \textbf{88.75}             & \textbf{79.78}                & \textbf{81.06}             & \textbf{70.37}                & \textbf{97.95/90.70}       & \textbf{95.98/83.02}      \\
\bottomrule[2pt]
\end{tabular}
}
\end{table*}

\begin{figure*}[t]
  \centering
  \includegraphics[width=1\linewidth]{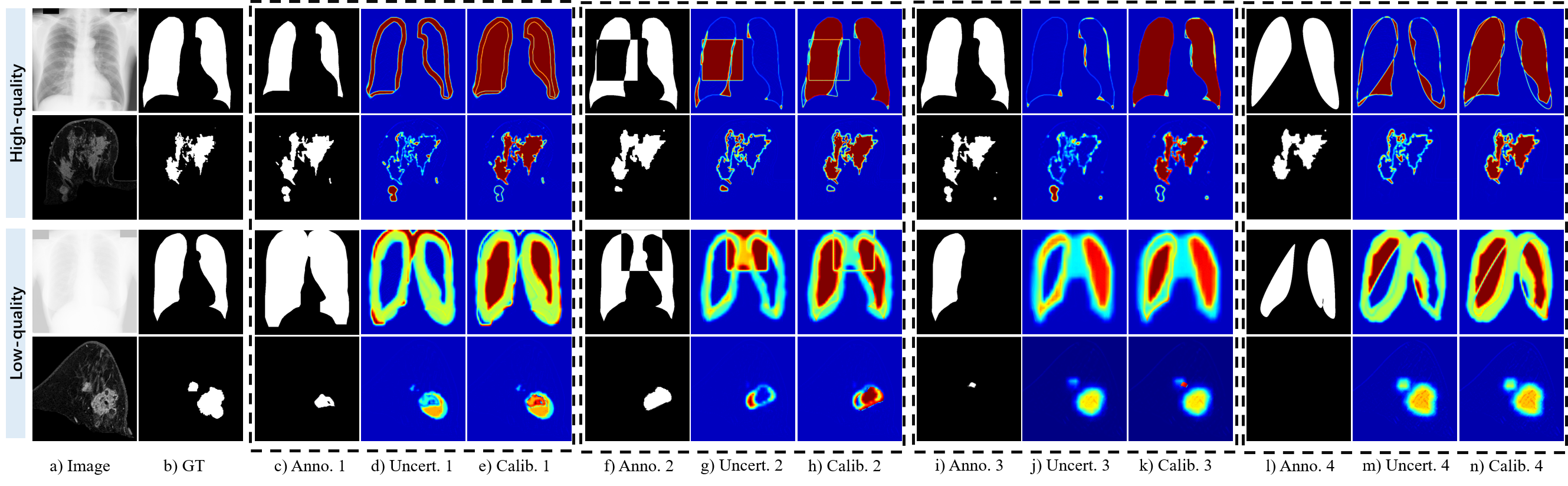}
  \caption{Illustration of images, their corresponding annotations, uncertainty maps, and calibration maps. The top two rows are high-quality samples, and the bottom two rows are low-quality samples. The uncertainty maps and calibration maps obtained by our UMA-Net clearly refelect the pixel-wise reliability of the annotations and the image quality of different samples.}\label{Analysis show}
\end{figure*}

\subsection{Analysis of the Method}

\subsubsection{Ablation Study}
We conducted detailed ablation experiments on these three datasets, which separately represent the multi-source annotation with simulated noise, rater-related noise, and systematic noise of handcraft algorithms. The results are shown in Table. \ref{Ablation result}. We use the training results of UNet as the baseline (BL), and the majority vote strategy generates the annotations for training the baseline UNet. We first added AUEM, which analyzed the commonality and difference among multiple annotations and generated uncertainty maps to weigh the loss of UNet. It can be seen that the metrics on the three datasets have been significantly increased, such as the dice score of the Lung segmentation dataset greatly grows from 75.28\% to 82.12\%. This shows that our AUEM is significantly better than the majority vote strategy. 
We then use consistency loss $L_c$, and the dice score of the Lung segmentation dataset continues to rise from 82.12\% to 85.29\%, which means consistency loss $L_c$ could strengthen AUEM's cross-referencing ability and improve the quality of the calibration maps and uncertainty maps. 
We further add QAM without the auxiliary predictor on this basis. When the samples are detected as low quality, the network parameters are not updated, which is equivalent to directly discarding low-quality data. The metrics of the three datasets increased, except for the optic cup, which decreased slightly. This is due to the RIGA dataset having more low-quality samples detected by QAM during the training stage, resulting in insufficient knowledge representation for the network to learn. This shows that learning from low-quality data does indeed degrade the performance of the network, and discarding them directly is not the best option, as this results in a loss of information. Finally, we added the auxiliary predictor to learn low-quality information separately, and the metrics of all three datasets showed growth. This shows that the auxiliary predictor can help the backbone learn useful representation information from low-quality samples. 
The result demonstrates that our method could effectively learn medical image segmentation from multi-source annotation datasets.

\subsubsection{Analysis of Uncertainty Maps}

We further provide a qualitative analysis of the uncertainty map from AUEM. Fig. \ref{Analysis show} illustrates four examples from two different datasets, each of them including the original image, ground truth, four annotations, and their corresponding uncertainty maps and calibration maps. The first two examples are identified as high-quality samples by QAM, while the last two examples are identified as low-quality samples. By observation, it can be noted that AUEM assigns large uncertainty values to unreliable pixel labels. Among the two high-quality samples, due to the high quality of the input image and the small discrepancy among annotations, their uncertainty maps are much more confident than low-quality samples. For the low-quality samples, because of the low image quality and large annotation variation, their uncertainty maps are underconfident, and calibration maps are in conflict. This shows that QAM can effectively identify low-quality samples and prevent them from directly undermining the network.


\subsubsection{Limitations and Discussion}
We acknowledge the limitation of the current work. 
First, AUEM requires all images to have the same amount of annotations, while a more feasible solution might allow adaptive adjustment to varied number of annotations.
Second, the current model feeds images of different qualities to different predictors, where a hard assignment strategy is performed here.
A potential improvement is using a dynamic assignment strategy such that a image could be learned by both predictors using quality-related weights. 
Our future work will delve into these challenges and foster the study on multi-source annotation-based segmentation.


\section{Conclusion}\label{conclusion}

In this paper, we propose UMA-Net to learn robust medical image segmentation from multi-source annotations. We develop an annotation uncertainty estimation module to analyze the similarities and differences among annotations and generate pixel-wise uncertainty maps to guide the segmentation network to learn from reliable pixels. To utilize the low-quality samples, we designed a quality assessment module to identify high-quality cases for segmentation learning. An additional auxiliary predictor of the segmentation network is also incorporated to preserve the maximum representation knowledge of low-quality samples. Through dual-level guidance, our method can effectively mitigate the side effects of noise. Extensive experiments on three datasets demonstrate the effectiveness and generalizability of our proposed method, showing promises in solving real-world clinical problems.

\bibliographystyle{IEEEtran}

\begin{thebibliography}{10}
\providecommand{\url}[1]{#1}
\csname url@samestyle\endcsname
\providecommand{\newblock}{\relax}
\providecommand{\bibinfo}[2]{#2}
\providecommand{\BIBentrySTDinterwordspacing}{\spaceskip=0pt\relax}
\providecommand{\BIBentryALTinterwordstretchfactor}{4}
\providecommand{\BIBentryALTinterwordspacing}{\spaceskip=\fontdimen2\font plus
\BIBentryALTinterwordstretchfactor\fontdimen3\font minus \fontdimen4\font\relax}
\providecommand{\BIBforeignlanguage}[2]{{%
\expandafter\ifx\csname l@#1\endcsname\relax
\typeout{** WARNING: IEEEtran.bst: No hyphenation pattern has been}%
\typeout{** loaded for the language `#1'. Using the pattern for}%
\typeout{** the default language instead.}%
\else
\language=\csname l@#1\endcsname
\fi
#2}}
\providecommand{\BIBdecl}{\relax}
\BIBdecl

\bibitem{litjens2017survey}
G.~Litjens, T.~Kooi, B.~E. Bejnordi, A.~A.~A. Setio, F.~Ciompi, M.~Ghafoorian, J.~A. Van Der~Laak, B.~Van~Ginneken, and C.~I. S{\'a}nchez, ``A survey on deep learning in medical image analysis,'' \emph{Medical image analysis}, vol.~42, pp. 60--88, 2017.

\bibitem{feng2022learning}
M.~Feng, K.~Liu, L.~Zhang, H.~Yu, Y.~Wang, and A.~Mian, ``Learning from pixel-level noisy label: A new perspective for light field saliency detection,'' in \emph{CVPR}, 2022, pp. 1756--1766.

\bibitem{almazroa2017agreement}
A.~Almazroa, S.~Alodhayb, E.~Osman, E.~Ramadan, M.~Hummadi, M.~Dlaim, M.~Alkatee, K.~Raahemifar, and V.~Lakshminarayanan, ``Agreement among ophthalmologists in marking the optic disc and optic cup in fundus images,'' \emph{International ophthalmology}, vol.~37, pp. 701--717, 2017.

\bibitem{orlando2020refuge}
J.~I. Orlando, H.~Fu, J.~B. Breda, K.~Van~Keer, D.~R. Bathula, A.~Diaz-Pinto, R.~Fang, P.-A. Heng, J.~Kim, J.~Lee \emph{et~al.}, ``Refuge challenge: A unified framework for evaluating automated methods for glaucoma assessment from fundus photographs,'' \emph{Medical image analysis}, vol.~59, p. 101570, 2020.

\bibitem{armato2011lung}
S.~G. Armato~III, G.~McLennan, L.~Bidaut, M.~F. McNitt-Gray, C.~R. Meyer, A.~P. Reeves, B.~Zhao, D.~R. Aberle, C.~I. Henschke, E.~A. Hoffman \emph{et~al.}, ``The lung image database consortium (lidc) and image database resource initiative (idri): a completed reference database of lung nodules on ct scans,'' \emph{Medical physics}, vol.~38, no.~2, pp. 915--931, 2011.

\bibitem{nguyen2019deepusps}
T.~Nguyen, M.~Dax, C.~K. Mummadi, N.~Ngo, T.~H.~P. Nguyen, Z.~Lou, and T.~Brox, ``Deepusps: Deep robust unsupervised saliency prediction via self-supervision,'' \emph{Advances in Neural Information Processing Systems}, vol.~32, 2019.

\bibitem{wang2022multi}
Y.~Wang, W.~Zhang, L.~Wang, T.~Liu, and H.~Lu, ``Multi-source uncertainty mining for deep unsupervised saliency detection,'' in \emph{CVPR}, 2022, pp. 11\,727--11\,736.

\bibitem{warfield2004simultaneous}
S.~K. Warfield, K.~H. Zou, and W.~M. Wells, ``Simultaneous truth and performance level estimation (staple): an algorithm for the validation of image segmentation,'' \emph{IEEE transactions on medical imaging}, vol.~23, no.~7, pp. 903--921, 2004.

\bibitem{ji2021learning}
W.~Ji, S.~Yu, J.~Wu, K.~Ma, C.~Bian, Q.~Bi, J.~Li, H.~Liu, L.~Cheng, and Y.~Zheng, ``Learning calibrated medical image segmentation via multi-rater agreement modeling,'' in \emph{CVPR}, 2021, pp. 12\,341--12\,351.

\bibitem{jensen2019improving}
M.~H. Jensen, D.~R. J{\o}rgensen, R.~Jalaboi, M.~E. Hansen, and M.~A. Olsen, ``Improving uncertainty estimation in convolutional neural networks using inter-rater agreement,'' in \emph{MICCAI 2019}.\hskip 1em plus 0.5em minus 0.4em\relax Springer, 2019, pp. 540--548.

\bibitem{jungo2018effect}
A.~Jungo, R.~Meier, E.~Ermis, M.~Blatti-Moreno, E.~Herrmann, R.~Wiest, and M.~Reyes, ``On the effect of inter-observer variability for a reliable estimation of uncertainty of medical image segmentation,'' in \emph{MICCAI 2018}.\hskip 1em plus 0.5em minus 0.4em\relax Springer, 2018, pp. 682--690.

\bibitem{wu2022learning}
J.~Wu, H.~Fang, Z.~Wang, D.~Yang, Y.~Yang, F.~Shang, W.~Zhou, and Y.~Xu, ``Learning self-calibrated optic disc and cup segmentation from multi-rater annotations,'' in \emph{MICCAI 2022}.\hskip 1em plus 0.5em minus 0.4em\relax Springer, 2022, pp. 614--624.

\bibitem{zhang2023multi}
F.~Zhang, Y.~Zheng, J.~Wu, X.~Yang, and X.~Che, ``Multi-rater label fusion based on an information bottleneck for fundus image segmentation,'' \emph{Biomedical Signal Processing and Control}, vol.~79, p. 104108, 2023.

\bibitem{ronneberger2015u}
O.~Ronneberger, P.~Fischer, and T.~Brox, ``U-net: Convolutional networks for biomedical image segmentation,'' in \emph{MICCAI 2015}.\hskip 1em plus 0.5em minus 0.4em\relax Springer, 2015, pp. 234--241.

\bibitem{9053405}
H.~Huang, L.~Lin, R.~Tong, H.~Hu, Q.~Zhang, Y.~Iwamoto, X.~Han, Y.-W. Chen, and J.~Wu, ``Unet 3+: A full-scale connected unet for medical image segmentation,'' in \emph{ICASSP 2020 - 2020}, 2020, pp. 1055--1059.

\bibitem{2020UNet}
Z.~Zhou, M.~M.~R. Siddiquee, N.~Tajbakhsh, and J.~Liang, ``Unet++: Redesigning skip connections to exploit multiscale features in image segmentation,'' \emph{IEEE Transactions on Medical Imaging}, vol.~39, no.~6, pp. 1856--1867, 2020.

\bibitem{huang2017densely}
G.~Huang, Z.~Liu, L.~Van Der~Maaten, and K.~Q. Weinberger, ``Densely connected convolutional networks,'' in \emph{Proceedings of the IEEE conference on computer vision and pattern recognition}, 2017, pp. 4700--4708.

\bibitem{li2018h}
X.~Li, H.~Chen, X.~Qi, Q.~Dou, C.-W. Fu, and P.-A. Heng, ``H-denseunet: hybrid densely connected unet for liver and tumor segmentation from ct volumes,'' \emph{IEEE transactions on medical imaging}, vol.~37, no.~12, pp. 2663--2674, 2018.

\bibitem{cciccek20163d}
{\"O}.~{\c{C}}i{\c{c}}ek, A.~Abdulkadir, S.~S. Lienkamp, T.~Brox, and O.~Ronneberger, ``3d u-net: learning dense volumetric segmentation from sparse annotation,'' in \emph{MICCAI 2016: 19th International Conference, Athens, Greece, October 17-21, 2016, Proceedings, Part II 19}.\hskip 1em plus 0.5em minus 0.4em\relax Springer, 2016, pp. 424--432.

\bibitem{karimi2020deep}
D.~Karimi, H.~Dou, S.~K. Warfield, and A.~Gholipour, ``Deep learning with noisy labels: Exploring techniques and remedies in medical image analysis,'' \emph{Medical image analysis}, vol.~65, p. 101759, 2020.

\bibitem{zhu2019pick}
H.~Zhu, J.~Shi, and J.~Wu, ``Pick-and-learn: Automatic quality evaluation for noisy-labeled image segmentation,'' in \emph{MICCAI 2019: 22nd International Conference, Shenzhen, China, October 13--17, 2019, Proceedings, Part VI 22}.\hskip 1em plus 0.5em minus 0.4em\relax Springer, 2019, pp. 576--584.

\bibitem{mirikharaji2019learning}
Z.~Mirikharaji, Y.~Yan, and G.~Hamarneh, ``Learning to segment skin lesions from noisy annotations,'' in \emph{First MICCAI Workshop, DART 2019, and First International Workshop, MIL3ID 2019, Shenzhen, Held in Conjunction with MICCAI 2019, Shenzhen, China, October 13 and 17, 2019, Proceedings 1}.\hskip 1em plus 0.5em minus 0.4em\relax Springer, 2019, pp. 207--215.

\bibitem{min2019two}
S.~Min, X.~Chen, Z.-J. Zha, F.~Wu, and Y.~Zhang, ``A two-stream mutual attention network for semi-supervised biomedical segmentation with noisy labels,'' in \emph{Proceedings of the AAAI Conference on Artificial Intelligence}, vol.~33, no.~01, 2019, pp. 4578--4585.

\bibitem{wang2020noise}
G.~Wang, X.~Liu, C.~Li, Z.~Xu, J.~Ruan, H.~Zhu, T.~Meng, K.~Li, N.~Huang, and S.~Zhang, ``A noise-robust framework for automatic segmentation of covid-19 pneumonia lesions from ct images,'' \emph{IEEE Transactions on Medical Imaging}, vol.~39, no.~8, pp. 2653--2663, 2020.

\bibitem{zhang2020robust}
T.~Zhang, L.~Yu, N.~Hu, S.~Lv, and S.~Gu, ``Robust medical image segmentation from non-expert annotations with tri-network,'' in \emph{MICCAI 2020}.\hskip 1em plus 0.5em minus 0.4em\relax Springer, 2020, pp. 249--258.

\bibitem{xu2022anti}
Z.~Xu, D.~Lu, J.~Luo, Y.~Wang, J.~Yan, K.~Ma, Y.~Zheng, and R.~K.-Y. Tong, ``Anti-interference from noisy labels: Mean-teacher-assisted confident learning for medical image segmentation,'' \emph{IEEE Transactions on Medical Imaging}, vol.~41, no.~11, pp. 3062--3073, 2022.

\bibitem{shi2021distilling}
J.~Shi and J.~Wu, ``Distilling effective supervision for robust medical image segmentation with noisy labels,'' in \emph{MICCAI 2021}.\hskip 1em plus 0.5em minus 0.4em\relax Springer, 2021, pp. 668--677.

\bibitem{wu2021uncertainty}
S.~Wu, C.~Chen, Z.~Xiong, X.~Chen, and X.~Sun, ``Uncertainty-aware label rectification for domain adaptive mitochondria segmentation,'' in \emph{MICCAI 2021}.\hskip 1em plus 0.5em minus 0.4em\relax Springer, 2021, pp. 191--200.

\bibitem{shiraishi2000development}
J.~Shiraishi, S.~Katsuragawa, J.~Ikezoe, T.~Matsumoto, T.~Kobayashi, K.-i. Komatsu, M.~Matsui, H.~Fujita, Y.~Kodera, and K.~Doi, ``Development of a digital image database for chest radiographs with and without a lung nodule: receiver operating characteristic analysis of radiologists' detection of pulmonary nodules,'' \emph{American Journal of Roentgenology}, vol. 174, no.~1, pp. 71--74, 2000.

\bibitem{saha2018machine}
A.~Saha, M.~R. Harowicz, L.~J. Grimm, C.~E. Kim, S.~V. Ghate, R.~Walsh, and M.~A. Mazurowski, ``A machine learning approach to radiogenomics of breast cancer: a study of 922 subjects and 529 dce-mri features,'' \emph{British journal of cancer}, vol. 119, no.~4, pp. 508--516, 2018.

\bibitem{isensee2021nnu}
F.~Isensee, P.~F. Jaeger, S.~A. Kohl, J.~Petersen, and K.~H. Maier-Hein, ``nnu-net: a self-configuring method for deep learning-based biomedical image segmentation,'' \emph{Nature methods}, vol.~18, no.~2, pp. 203--211, 2021.

\bibitem{zhang2019attention}
S.~Zhang, H.~Fu, Y.~Yan, Y.~Zhang, Q.~Wu, M.~Yang, M.~Tan, and Y.~Xu, ``Attention guided network for retinal image segmentation,'' in \emph{MICCAI 2019}.\hskip 1em plus 0.5em minus 0.4em\relax Springer, 2019, pp. 797--805.

\bibitem{wang2019boundary}
S.~Wang, L.~Yu, K.~Li, X.~Yang, C.-W. Fu, and P.-A. Heng, ``Boundary and entropy-driven adversarial learning for fundus image segmentation,'' in \emph{MICCAI 2019}.\hskip 1em plus 0.5em minus 0.4em\relax Springer, 2019, pp. 102--110.

\bibitem{luo2023deep}
L.~Luo, X.~Wang, Y.~Lin, X.~Ma, A.~Tan, R.~Chan, V.~Vardhanabhuti, W.~C. Chu, K.-T. Cheng, and H.~Chen, ``Deep learning in breast cancer imaging: A decade of progress and future directions,'' 2023.

\bibitem{zhou2019weakly}
J.~Zhou, L.-Y. Luo, Q.~Dou, H.~Chen, C.~Chen, G.-J. Li, Z.-F. Jiang, and P.-A. Heng, ``Weakly supervised 3d deep learning for breast cancer classification and localization of the lesions in mr images,'' \emph{Journal of Magnetic Resonance Imaging}, vol.~50, no.~4, pp. 1144--1151, 2019.

\bibitem{luo2019deep}
L.~Luo, H.~Chen, X.~Wang, Q.~Dou, H.~Lin, J.~Zhou, G.~Li, and P.-A. Heng, ``Deep angular embedding and feature correlation attention for breast mri cancer analysis,'' in \emph{MICCAI 2019}.\hskip 1em plus 0.5em minus 0.4em\relax Springer, 2019, pp. 504--512.

\bibitem{wei2018three}
D.~Wei, S.~Weinstein, M.-K. Hsieh, L.~Pantalone, and D.~Kontos, ``Three-dimensional whole breast segmentation in sagittal and axial breast mri with dense depth field modeling and localized self-adaptation for chest-wall line detection,'' \emph{IEEE Transactions on Biomedical Engineering}, vol.~66, no.~6, pp. 1567--1579, 2018.

\bibitem{militello2021unsupervised}
C.~Militello, A.~Ranieri, L.~Rundo, I.~D’Angelo, F.~Marinozzi, T.~V. Bartolotta, F.~Bini, and G.~Russo, ``On unsupervised methods for medical image segmentation: Investigating classic approaches in breast cancer dce-mri,'' \emph{Applied Sciences}, vol.~12, no.~1, p. 162, 2021.

\end{thebibliography}

\end{document}